\documentstyle[preprint,aps]{revtex}

\begin{document}
\draft

\title{Experimental search for solar axions via coherent
Primakoff conversion in a germanium spectrometer}
%CONVERSION IN A GERMANIUM SPECTROMETER}

\author{F.T. Avignone III$^a$. D. Abriola$^b$, R.L. Brodzinski$^c$,
J.I. Collar$^d$, R.J. Creswick$^a$, \\
D.E. DiGregorio$^b$, H.A. Farach$^a$, A.O. Gattone$^b$, C.K. Gu\'erard$^{a,b}$,
F. Hasenbalg$^b$, H. Huck$^b$, \\
H.S. Miley$^c$, A. Morales$^e$, J. Morales$^e$, S. Nussinov$^f$,
A. Ortiz de Sol\'orzano$^e$, J.H. Reeves$^c$, J.A. Villar$^e$, and
K. Zioutas$^g$ (The SOLAX Collaboration)}

\address{$^a$ Department of Physics and Astronomy, University of South
Carolina, Columbia, SC 29208 USA \\
$^b$ Department of Physics, TANDAR Laboratory, C.N.E.A., Buenos Aires,
Argentina \\
$^c$ Pacific Northwest National Laboratory, Richland, WA 99352 USA \\
$^d$ CERN, CH-1211 Geneva, 23 Switzerland \\
$^e$ Laboratorio de F\'{\i}sica Nuclear y Altas Energias, Universidad
de Zaragoza, Zaragoza, Spain \\
$^f$ Department of Physics, Tel Aviv University, Tel Aviv, Israel \\
$^g$ Department of Physics, University of Thessaloniki, GR54006 Thessaloniki,
Greece}

\date{}
\maketitle

\begin{abstract}
Results are reported of an experimental search for the unique, rapidly varying
temporal pattern of solar axions coherently converting into photons via the
Primakoff effect in a single crystal germanium detector.  This conversion is
predicted when axions are incident at a Bragg angle with a crystalline plane.
The analysis of approximately 1.94~kg.yr of data from the 1~kg DEMOS detector
in Sierra Grande, Argentina yields a new laboratory bound on axion--photon
coupling of $g_{a\gamma\gamma} < 2.7 \times 10^{-9}$~GeV$^{-1}$, independent
of axion mass up to $\sim 1$~keV.
\end{abstract}

\pacs{}

\vspace{1cm}

Early QCD theories predicted a particle with the quantum numbers of
the $\eta$--meson, but with a mass close to that of the pion
\cite{[1]}.  A term added to the QCD Lagrangian to ameliorate this
so--called U(1) problem violated CP invariance in strong interactions
and implied a neutron electric--dipole moment about 10$^9$ times larger
than the experimental upper bound \cite{[2]}.  Peccei and Quinn
\cite{[3]} introduced a new field causing strong CP--violation to vanish
dynamically.  Subsequently, Weinberg \cite{[4]} and Wilczek \cite{[5]}
demonstrated that the  Peccei--Quinn mechanism generates a
Nambu--Goldstone boson, the axion, that mixes with the $\pi^{\rm o}$ to
acquire a small mass.

Extensive reviews of axion phenomenology, and their effects on stellar
evolution, have been given by Raffelt \cite{[6],[7]}.  A detailed
treatment of solar axions and of a proposed method of detecting them
was given by van Bibber, et al. \cite{[8]}.  Details of a theory for
searching for axions with germanium detectors was recently given by
Creswick, et al. \cite{[9]} and will not be repeated here.  The
objective of this experiment is to detect solar axions through their
coherent Primakoff conversion (see Figure 1) into photons in the lattice of a
germanium crystal when the incident angle satisfies the Bragg
condition.  As shown in \cite{[9]}, the detection rates in various
energy windows are correlated
with the relative orientations of the detector and the sun.  This
correlation results in a temporal structure which should be a
distinctive, unique signature of the axion.  In this letter, the
results of a search using a 1~kg, ultra--low background germanium
detector installed in the HIPARSA iron mine in Sierra Grande, Argentina
at 41$^{\rm o}$~41'~24''~S and 65$^{\rm o}$~22'~W are presented.  A
complete description of the experimental set--up was given earlier by
Di Gregorio, et al.\cite{[10]} and Abriola, et al. \cite{[11]}.  This
experiment was motivated by earlier papers by Buchm\"uller and
Hoogeveen \cite{[12]} and by Paschos and  Zioutas \cite{[13]}; the
present technique was originally suggested by Zioutas and developed by
Creswick, et al. \cite{[9]}.

The vertical axis of our detector is the (100) crystalline axis.  The
orientation of the (010) and (001) axes are unknown at this time.
Therefore, to place a bound on the axion interaction rate, the data
must be analyzed for many azimuthal orientations of the crystal, and
the weakest bound selected.  To confirm a positive effect, either the
azimuthal orientation of the (010) and (001) axes must be known or the
experiment must be repeated with the crystal at two or more azimuthal
orientations which would have very different calculated responses.

The  terrestrial flux of axions from the sun can be approximated by the
expression \cite{[8],[9]}:

\begin{equation}
  \frac{d\Phi}{dE}\;=\; \lambda^{1/2} \,
  \frac{\Phi_{\rm_o}(E/E_{\rm o})^3}{E_{\rm_o}(e^{E/E_{\rm o}}\,-\,1)} \; ,
\end{equation}
where $\lambda = (g_{a\gamma \gamma} \times 10^8)^4$ and is
dimensionless, $E_{\rm o} = 1.103$~keV, and $\Phi_{\rm o} = 5.95 \times
10^{14}$~cm$^{-2}$~sec$^{-1}$. The total flux for $\lambda = 1$
integrated from 0 to 12~keV is $3.54 \times
10^{15}$~cm$^{-2}$~sec$^{-1}$. The spectrum is a continuum peaking at
about 4~keV decreasing to a negligible contribution above 8~keV. The
differential
cross section for Primakoff conversion on an atom with nuclear charge
$Ze$ is \cite{[9]}:

\begin{equation}
    \frac{d\sigma}{d\Omega} \;=\; \left [
    \frac{Z^2 \alpha \hbar^2 c^2 g_{a\gamma \gamma}^2}{16 \pi} \right] \,
    \frac{q^2\,(4k^2 \,-\, q^2)}{(q^2\,+\,r_{\rm o}^{-2})^2} \; ,
\end{equation}
where $q$ is the momentum transfer, $k$ is the momentum of the incoming
axion, and $r_{\rm o}$ is the screening length of the atom
in the lattice.  For germanium, $\sigma_{\rm o} = Z^2 \alpha \hbar^2 c^2
g_{a\gamma \gamma}/8 \pi = 1.15 \times 10^{-44}$~cm$^2$ when
$g_{a\gamma \gamma} = 10^{-8}$~GeV$^{-1}$, or equivalently $\lambda =
1$.

For light axions the Primakoff process in a periodic lattice is
coherent when the Bragg condition ($2 d \sin\theta = n\lambda$)
is satisfied, that is when $\vec{q}$ transferred to the crystal is a
reciprocal lattice vector $\vec{G} = 2\pi(h,k,l)/a_{\rm o}$. Here
$a_{\rm o}$ is the size of the conventional cubic cell, and $h$, $k$,
and $l$ are integers \cite{[14]}.

It was shown that the rate of conversion of axions with energy $E$ when
the sun is in the direction $\hat{k}$, $\dot{N}(\hat{k},E)$, can be
written \cite{[9]}:

\begin{equation}
  \dot{N}(\hat{k},E) \;=\; 2\hbar c \,\frac{V}{v_c}\, \sum_G \left| S(G)
  \right|^2 \,
   \frac{d\sigma}{d\Omega}(\vec{G})\, \frac{1}{|\vec{G}|^2}\,
     \frac{d\Phi}{dE}
      \, \delta(E-\frac{\hbar c \left| \vec{G}\right|^2}
                       {2\hat{k}\cdot \vec{G}}) \; ,
\end{equation}
where $V$ is the volume of the crystal, $v_c$ is the volume of a unit
cell, $S(G)$ is the structure function for germanium, and $d\Phi/dE$ is
evaluated at the axion energy of $\hbar c |\vec{G}|^2/ 2\hat{k} \cdot
\vec{G}$. The structure function for germanium is:

\begin{equation}
  S(G)\; =\; \left[ 1 \,+\, e^{i\pi(h+k+l)/2} \right] \,
  \left[ 1\,+\,e^{i\pi(h+k)} \,+\, e^{i\pi(h+l)} \,+\, e^{i\pi(k+l)}
  \right] \; .
\end{equation}

Note that in (3) the coherent conversion of axions occurs only for a
particular axion energy given the position of the sun, $\hat{k}$, and
reciprocal lattice vector $\vec{G}$. However, the detector has a finite
energy resolution; for the detector in Sierra Grande it is 1~keV FWHM
at 10~keV. We take this into account by smoothing $\dot{N}(\hat{k},E)$
with a Gaussian of the appropriate width. Finally, we take the relevant
part of the energy spectrum, in this case from the threshold energy of
4~keV up to 8~keV (which is just below the X-rays at 10~keV), and
calculate the total rate of conversion in windows of width $\Delta E$,
typically 0.5~keV,

\begin{eqnarray}
 R(\hat{k},E)\; =\; 2\hbar c\,\frac{V}{v_c}\, \sum_G \left| S(G) \right|^2 \,
   \frac{d\sigma}{d\Omega} \, \frac{1}{ \left| \vec{G}\right|^2} \,
   \frac{d\Phi}{dE} \, \frac{1}{2} \,\nonumber \\
   \left[ erf\left(\frac{E-E_a(\hat{k},\vec{G})}
   {\sqrt{2}\sigma}\right) - erf
   \left(\frac{E-E_a(\hat{k},\vec{G})-\Delta E}
   {\sqrt{2}\sigma}\right)\right],
\end{eqnarray}
where $E_a(\hat{k},G)=\frac{\hbar c \left| G \right|^2}
{2\hat{k}\cdot \vec{G}}$ and $erf(x)=\frac{2}{\sqrt{\pi}}\int_0^x e^{-t^2}dt$
is the error function. In equation
(5) we have neglected the angular size of the core of the sun and
the mass of the axion which is justified when
$m_a c^2$ is small compared to the core temperature of the sun
\cite{[12]}, i.e., up to a few keV.

The theoretical axion
detection rate for this detector, calculated with equation (5), is
shown in Figure 2.  The position of the sun is computed at any instant
in time using the U.S. Naval Observatory Subroutines (NOVAS)
\cite{[15]}.  The pronounced variation in $R(\hat{k},E)$ as a function
of time invites the data to be analyzed with the correlation function:

\begin{equation}
   \chi\; \equiv \; \sum_{i=1}^n\, \left[
       R(t_i,E)\; -\, <R(E)> \right] \, n(t_i) \; ,
\end{equation}
where $R(t_i,E)$ is the smooth shape of the theoretical rate at the
instant of time, $t_i$, $<R(E)>$ is the average rate over a finite time
interval, and $n(t_i)$ is the number of events at $t_i$ in a time
interval $\Delta t$, usually 0 or
1. The choice for the weighting function $W(t,E)=R\left(\hat{k}(t),E \right)
- \left<R(E)\right>$ is motivated by the requirement that any constant
background average to zero in $\chi$, whereas a counting rate which follows
$R\left( \hat{k}(t),E \right)$ increases $\chi$.

The number of counts at time, $t$, in the interval $\Delta t$
is assumed to be due in part to axions and in part to background
governed by a Poisson process with mean:

\begin{equation}
  <n(t)>\; =\; \left[ \lambda\, R(t,E)\, +\,b(E) \right]\, \Delta t,
\end{equation}
where $b(E)$ is constant in time.

The average value of $\chi$ is then,

\begin{eqnarray}
 <\chi> & \;=\; & \sum_i\, \left[ R(t_i,E) \; -\, <R(E)> \right]\,
   \left[ \lambda\, R(t_i,E)\, +\,b(E) \right]\, \Delta t  \nonumber \\
   & \; =\; & \sum_i\, W(t_i,E)\,
   \left[ \lambda\, R(t_i,E)\, +\,b(E) \right]\, \Delta t .
\end{eqnarray}

We can add and subtract the constant quantity $\lambda <R(E)>$ to the
second factor in equation (8). Any time independent contributions
multiplied by $W(t,E)$ in eq. (8), and summed over time, will vanish.
Accordingly, taking the limit as $\Delta t \rightarrow 0$, we obtain:

\begin{equation}
   <\chi(\lambda)>\; =\; \lambda \, \int_0^T\, W^2(t,E)\, dt.
\end{equation}
The expected uncertainty in $\chi$, $(\Delta \chi)^2 = <\chi^2> -
<\chi>^2$, is given by,

\begin{eqnarray}
\lefteqn{\hspace{-3.5cm} (\Delta \chi)^2 \;=\; \sum_i\, \sum_j \,
W(t_i,E) W(t_j,E)\, \left[ <n(t_i) n(t_j)>\, -\,
  <n(t_i)> <n(t_j)> \right]\; = }  \nonumber \\
 & \; = \; & \sum_i \, W^2(t_i,E) \,
  \left[ <n(t_i)^2> \, -\, <n(t_i)>^2 \right],
\end{eqnarray}
where the square bracket is $<\Delta n(t_i)>^2$, which in Poisson
statistics is $<n(t_i)>$. Accordingly,

\begin{equation}
   (\Delta \chi)^2 \;=\; \sum_{i}\, W^2(t_i,E) <n(t_i)>.
\end{equation}
By (7) we have:

\begin{eqnarray}
  (\Delta \chi)^2  &\;=\; & \sum_{i}\, W^2(t_i,E)\, \left[ \lambda\,
  R(t_i,E)\, +\,b(E) \right]\, \Delta t,  \nonumber \\
  & \; =\; &
  \sum_{i}\, W^2(i,E)\, \left\{ \lambda \left[ R(t_i,E) \; -\,
  <R(E)> \right]\, +\, \lambda <R(E)>\, + \;b(E) \right\} \, \Delta t,
\end{eqnarray}
which in the limit $\Delta t \rightarrow 0$ becomes,

\begin{equation}
   (\Delta \chi)^2 \;=\; \lambda \, \int_0^T\, W^3(t,E)\, dt \;+\;
      R_T(E) \, \int_0^T\, W^2(t,E)\, dt.
\end{equation}
The quantity $ R_T(t,E)$ is the average total counting rate, including both
axion conversions and background.

The data are separately analyzed in energy bins, $\Delta E_k$, fixed by
the detector resolution (FWHM $\sim 1$~keV in this case). The
likelihood function is then constructed:

\begin{equation}
L(\lambda) \;=\; \prod_k\, \exp \left[
   \frac{-(\chi_k \,-\, <\chi_k>)^2}{2 (\Delta \chi_k)^2} \right].
\end{equation}
To an excellent approximation $(\Delta \chi_k)^2$ is dominated by
background. Maximizing the likelihood function, the most probable
value of $\lambda$ is given by

\begin{equation}
  \lambda_{\rm o}\; =\; \sum_k \, \chi_k \,/\, \sum_k\, A_k,
\end{equation}
where,
\begin{equation}
  A_k\; \equiv\; \int_0^T\, W_k^2(t,E)\, dt ,
\end{equation}
and the width of the likelihood function is given by,
\begin{equation}
  \sigma_{\lambda} \;=\; \left( \sum_k \, A_k\,/\, b_k \right)^{-1/2}.
\end{equation}
We note that $A_k$ is proportional to the time of the experiment, so
that $\sigma_{\lambda}$ decreases as $T^{1/2}$. The background scales
with the detector mass, while $A_k$ scales as the square of the
detector mass, therefore $\sigma_{\lambda}$ decreases as $(M_d
T)^{-1/2}$.

We have carried out extensive Monte-Carlo test of this method of
analysis. As a test of our analysis, typical results for the likelihood
function
for the cases $\lambda = 0$ (no axions) and $\lambda = 0.003$
were calculated
with realistic backgrounds for a detector operating with the same
mass, energy resolution, and threshold as the DEMOS detector at the
latitude and longitude of Sierra Grande for one year. It is clear from
this calculation that
the correlation function analysis is consistent and quite sensitive to
the presence of a variation in the counting range due to solar
axions with a signal to noise ratio
less than 1\%.

A scatter plot of the distribution of events on the celestial sphere
is shown in Figure 3. The variation in the density of events reflects
the amount of time the sun spends at each position in the sky, and is
consistent with no effect.

As mentioned above, the azimuthal orientation of the germanium
crystal is not known, which requires the analysis be carried out as a
function of the angle $\phi$, defined as the angle between the (010)
axis and true north. The results of these calculations for 707~days of
data in the energy range from 4 to 8~keV in 0.5~keV intervals are shown
in Figure 4.

The most conservative upper bound on $\lambda$, or equivalently $g_{a
\gamma \gamma}$, is found by taking for $\phi$ the angle at which
$\lambda$ is maximum. This yields an upper bound on the axion--photon
coupling constant $g_{a \gamma \gamma} < 2.7 \times 10^{-9}$~GeV$^{-1}$
at the 95\% confidence level.

In Figure 5 we show the area of the axion mass -- coupling constant
plane excluded by this result along with results of earlier work.

While this bound is interesting because it is a laboratory constraint,
it does not challenge the bound placed by Raffelt \cite{[7]}, $g_{a
\gamma \gamma} \leq 10^{-10}$~GeV$^{-1}$ based on the helium burning
rate in low mass stars. A coupling constant $g_{a \gamma \gamma} \simeq
10^{-9}$~GeV$^{-1}$ would imply axion emission rates 100 times higher
than the stellar bound, and a significantly different concept of
stellar evolution.

This experiment can be considerably improved by using a large number
of smaller p--type germanium detectors with known orientations of the
(010) axes, with energy thresholds below 2~keV, and energy
resolutions corresponding to FWHM $\approx$ 0.5~keV. This is being
proposed at this time. Another collaboration could also operate the
COSME experiment in the new University of Zaragoza underground
laboratory in the Canfranc tunnel at 42$^{\rm o}$ 48'~N and 0$^{\rm o}$
31'~W.  The COSME detector is a 0.25~kg crystal having an energy
threshold of $\sim$1.8 keV and a resolution of 0.5~keV FWHM at
10~keV.   A positive result in this northern hemisphere experiment over
a wider energy range should have a very different temporal pattern from
that of the Sierra Grande experiment, but should yield the same value
of $\lambda$.

One of the USC/PNNL twin detectors is currently operating in the Baksan
Neutrino Observatory in Russia at 660~mwe, and if moved to a location
with greater overburden could also be used to acquire meaningful data
on solar axion rates.

The analysis presented here allows one to legitimately combine the
results of a number of experiments. Accordingly, the results from a
large number of experiments located throughout the world can be
combined to yield results equivalent to a single large experiment.

\newpage

\begin{center}
ACKNOWLEDGEMENTS
\end{center}

This work was supported by the U.S. National Science Foundation (NSF)
under grant INT930INT1522, the U.S. Department of Energy (DOE) under
contract DE--AC06-- 76RLO 1830, the Consejo Nacional de Investigaciones
Cient\'{\i}ficas y T\'ecnicas (CONICET) and Fundacion Antorchas of
Argentina, and the Spanish Agency for Science and Technology (CICYT)
under grant AEN96--1657.  We also thank the personnel of the HIPARSA
iron mine for significant assistance during installation of the
experimental equipment and J. A. Bangert of the U. S. Naval Observatory
for supplying their vector astronomy subroutines.

\begin{figure}
\caption{Feynman diagram of the Primakoff conversion of axions into photons.}
\end{figure}

\begin{figure}
\caption{A typical axion--photon conversion rate, $R(t,E)$, for various
energy bands. The experimental energy resolution FWHM = 1.0~keV at
10~keV was used.}
\end{figure}

\begin{figure}
\caption{Scatter plot showing the distribution on the celestial sphere
of all events between 4~keV and 8~keV over the data collection period
of 707 days.}
\end{figure}

\begin{figure}
\caption{Values of $\lambda$ calculated from the 707 days of data as a
function of the azimuthal angle $\phi$. The error bars are $1 \sigma$.}
\end{figure}

\begin{figure}
\caption{Exclusion plots on the $g_{a \gamma \gamma}$ vs. axion--mass
plane.  The curve to the left are from Ref.~[16]. The letters indicate
the following helium gas pressures in the conversion region of a strong
magnetic field:  (a) vacuum, (b) 55~Torr, and (c) 100 Torr.}
\end{figure}


\begin{references}

\bibitem{[1]} S. Weinberg, Phys. Rev. {\bf D11} (1975) 3583;  S. Weinberg,
``The Quantum Theory of Fields'', Vol. II, Cambridge Press, New York, NY
1996 pp 243.

\bibitem{[2]} N. F. Ramsey, Rep. Prog. Phys. {\bf 45} (1982) 95;  Also see
Review of Particle Properties, Phys. Rev. D (1994) 1218.

\bibitem{[3]} R. D. Peccei and H. Quinn, Phys. Rev. Lett. {\bf 38} (1977)
1440;  Phys. Rev. {\bf Dl6} (1977) 1791.

\bibitem{[4]} S. Weinberg, Phys. Rev. Lett. {\bf 40} (1978) 223.

\bibitem{[5]} F. Wilczek, Phys. Rev. Lett. {\bf 40} (1978) 279.

\bibitem{[6]} G. G. Raffelt, ``Stars as Laboratories for Fundamental Physics'',
 University of Chicago Press, Chicago, 1996.

\bibitem{[7]} G. G. Raffelt, Phys. Reports {\bf 198} (1990) 1.

\bibitem{[8]} K. van Bibber, P. M. McIntyre, D. E. Morris, G. G. Raffelt,
Phys. Rev. {\bf D39} (1989) 2089.

\bibitem{[9]} R. J. Creswick, F. T. Avignone III, H. A. Farach, J. I.
Collar, A. O. Gattone, S. Nussinov, and K. Zioutas, (Submitted to Phys.
Lett. 1997).

\bibitem{[10]} D. E. Di Gregorio, D. Abriola, F. T. Avignone III, R. L.
Brodzinski, J. I. Collar, H. A. Farach, E. Garc\'{\i}a, A. O. Gattone, F.
Hasenbalg, H. Huck, H. S. Miley, A. Morales, J. Morales, A. Ortiz de
Sol\'orzano, J. Puimed\'on, J. H. Reeves, C. S\'aenz, A. Salinas, M. L.
Sarsa, D. Tomasi, I. Urteaga, and J. A. Villar, Nucl. Phys. B (Proc.
Suppl.) {\bf 48} (1996) 56

\bibitem{[11]} D. Abriola, F. T. Avignone III, R. L. Brodzinski, J. I.
Collar, D. E. Di Gregorio, H. A. Farach, E. Garc\'{\i}a, A. O. Gattone, F.
Hasenbalg, H. Huck, H. S. Miley, A. Morales, J. Morales, A. Ortiz de
Sol\'orzano, J. Puimed\'on, J. H. Reeves, C. S\'aenz, A. Salinas, M. L.
Sarsa, D. Tomasi, I. Urteaga, and J. A. Villar, Astropart. Phys. {\bf 6}
(1996) 63.

\bibitem{[12]} W. Buchm\"uller and F. Hoogeveen, Phys. Lett. {\bf B237}
(1990) 278.

\bibitem{[13]} E. A. Paschos and K. Zioutas, Phys. Lett. {\bf B323} (1994) 367.

\bibitem{[14]} Charles Kittel, ``Introduction to Solid State Physics'',
Sixth Edition, John Wiley \&
Sons, Inc. New York (1986) pp29.

\bibitem{[15]} G. H. Kaplan, J. A. Huges, P. K. Seidelmann, C. A. Smith,
and B. D. Yallop,
Astronomical Journal {\bf 97} (1989) 1197.

\bibitem{[16]} D. M. Lazarus, G. C. Smith, R. Cameron, A. C. Melissinos, G.
Ruoso, Y. K. Semertzidis, and F. A. Nezrick. Phys. Rev. Lett. {\bf 69}
(1992) 2333.

\end{references}
\end{document}